\documentclass[preprint,
bibnotes,amsmath,amssymb,aps,prb,superscriptaddress,unicode-math]{revtex4-2}
\DeclareUnicodeCharacter{2212}{\ensuremath{-}}

\usepackage{graphicx}
\usepackage{dcolumn}
\usepackage{bm}
\usepackage[utf8]{inputenc}
\usepackage{caption}
\begin{document}

\preprint{APS/123-QED}

\title{Tuning exciton emission via ferroelectric polarization at a heterogeneous interface between a monolayer transition metal dichalcogenide and a perovskite oxide membrane}

\author{Jaehong Choi}
\email{jc3452@cornell.edu}
\affiliation{School of Applied and Engineering Physics, Cornell University, Ithaca, NY 14850, USA}

\author{Kevin J. Crust}%
\affiliation{Department of Physics, Stanford University, Stanford, CA 94305, USA}
 \affiliation{Stanford Institute for Materials and Energy Sciences,
SLAC National Accelerator Laboratory, Menlo Park, CA 94025, USA}
 
\author{Lizhong Li}
\affiliation{School of Applied and Engineering Physics, Cornell University, Ithaca, NY 14850, USA}

\author{Kihong Lee}
\affiliation{Department of Physics, Cornell University, Ithaca, NY 14850, USA}

\author{Jialun Luo}
\affiliation{Department of Physics, Cornell University, Ithaca, NY 14850, USA}

\author{Jae-Pil So}
\affiliation{School of Applied and Engineering Physics, Cornell University, Ithaca, NY 14850, USA}

\author{Kenji Watanabe}
\affiliation{Research Center for Electronic and Optical Materials, National Institute for Materials Science, 1-1 Namiki, Tsukuba 305-0044, Japan}

\author{Takashi Taniguchi}
\affiliation{Research Center for Electronic and Optical Materials, National Institute for Materials Science, 1-1 Namiki, Tsukuba 305-0044, Japan}

\author{Harold Y. Hwang}
\affiliation{Stanford Institute for Materials and Energy Sciences,
SLAC National Accelerator Laboratory, Menlo Park, CA 94025, USA}
 \affiliation{Department of Applied Physics, Stanford University, Stanford, CA 94305, USA}

\author{Kin Fai Mak}
\affiliation{School of Applied and Engineering Physics, Cornell University, Ithaca, NY 14850, USA}
 \affiliation{Department of Physics, Cornell University, Ithaca, NY 14850, USA}
  \affiliation{Kavli Institute at Cornell for Nanoscale Science, Ithaca, NY 14850, USA}

\author{Jie Shan}
\affiliation{School of Applied and Engineering Physics, Cornell University, Ithaca, NY 14850, USA}
 \affiliation{Department of Physics, Cornell University, Ithaca, NY 14850, USA}
  \affiliation{Kavli Institute at Cornell for Nanoscale Science, Ithaca, NY 14850, USA}

\author{Gregory D. Fuchs}
\email{gdf9@cornell.edu}
\affiliation{School of Applied and Engineering Physics, Cornell University, Ithaca, NY 14850, USA}
\affiliation{Kavli Institute at Cornell for Nanoscale Science, Ithaca, NY 14850, USA}

\begin{abstract}
We demonstrate the integration of a thin BaTiO$_3$ (BTO) membrane with monolayer MoSe$_2$ in a dual gate device that enables in-situ manipulation of the BTO ferroelectric polarization with a voltage pulse. While two-dimensional (2D) transition metal dichalcogenides (TMDs) offer remarkable adaptability, their hybrid integration with other families of functional materials beyond the realm of 2D materials has been challenging. Released functional oxide membranes offer a solution for 2D/3D integration via stacking. 2D TMD excitons can serve as a local probe of the ferroelectric polarization in BTO at a heterogeneous interface. Using photoluminescence (PL) of MoSe$_2$ excitons to optically readout the doping level, we find that the relative population of charge carriers in MoSe$_2$ depends sensitively on the ferroelectric polarization. This finding points to a promising avenue for future-generations versatile sensing devices with high sensitivity, fast read-out, and diverse applicability for advanced signal processing. 

\end{abstract}
\maketitle


\section{\label{sec:level1} Introduction}
Heterogeneous 2D-3D hybrid interfaces provide an interesting platform to study the properties of functional materials. The ultrathin 2D TMDs offer remarkable flexibility for integration and sensitivity to external perturbations. Compared to their bulk counterparts, dielectric screening is reduced in 2D and this increases the Coulomb interaction, leading to strong exciton physics. The direct band gap of 2D TMDs results in excitons dominating the optical response, enabling their applications to gas, chemical, and biosensing \cite{Lee2018,Kim2024,Rohaizad2021,Bolosky2019,Anichini2018, Lemme2022}. The sensing mechanism relies on variations in both optical and electrical properties in the presence of gases, ions, or biomolecule. Utilizing a fluorescence probe offers distinct advantages. In contrast to electronic sensors that permit only global detection, excitons in 2D TMDs enable local sensing with rapid read-out and high sensitivity, capable of capturing fast dynamic phenomena \cite{Feierabend2017}.

The van der Waals (vdW) interaction simplifies integration of 2D TMDs with diverse functional materials, such as ferroelectric materials, eliminating the need for lattice matching. Using a straightforward `tear-and-stack' technique, 2D ferroelectric materials like  CuInP$_2$S$_6$ (CIPS) and In$_2$Se$_3$ \cite{Ping2022, Mao2021, Si2018, Jiang2021, Huo2022} can be mechanically exfoliated and stacked in various configurations, providing a versatile exploration of geometries and highly sensitive ferroelectric polarization detection. The integration also extends beyond 2D ferroelectric materials, encompassing an organic ferroelectric P(VDF-TrFe) \cite{Zhu2022, Wu2020, Lv2019, Wang2015}, bulk perovskite oxides such as LiNbO$_{3}$ \cite{Wen2019, Nguyen2015}, and many others \cite{Sun2017, Chen2017, Salazar2022, Tao2017, Li2018, Chen2020}. However, the application of the `tear-and-stack' technique faces limitations for 3D materials, including bulk perovskite oxides, due to inherent structural constraints.

Among non-2D ferroelectric materials, perovskite oxides have strong ferroelectricity and a large dielectric constant. Perovskite oxides integrated with 2D TMDs to study doping and electronic structures at the heterogeneous interface were often in bulk forms \cite{Chen2017, Salazar2022, Tao2017, Li2018}. The integration of perovskite oxide thin films with 2D TMDs is challenging because of the epitaxial methods necessary for obtaining high quality material synthesis. Thin perovskite oxides are grown on a limited set of substrates, which has constrained the range of device structures that can be explored \cite{Chiabrera2022, Kum2020}. Recent studies have shown that using water-soluble sacrificial buffer layers allow the separation of highly crystalline perovskite oxide membranes from the growth substrates without inducing structural damage \cite{Lu2016, Lu2019, Dong2020, Baek2017, Seung2020, Davidovikj2020, Harbola2021, Harbola2021-2}. This approach enables flexible integration of oxide membranes in a broad spectrum of device geometries, extending the scope of potential applications and science opportunities. 

Recent work has demonstrated the vertical stacking of perovskite oxide membranes with 2D TMDs in a field effect transistor (FET) \cite{Yang2022, Huang2022, Puebla2022}. Yang et al.~and Huang et al.~used a released SrTiO$_3$ (STO) membrane as a dielectric layer in 2D TMD-based FETs. They exploited the large dielectric constant of STO to create a transistor with a large ON/OFF ratio (10$^8$) \cite{Yang2022} and low leakage current arising from the sub-one-nanometer capacitance equivalent thickness of the STO membrane \cite{Huang2022}. Puebla et al.~used an isolated BaTiO$_3$ (BTO) membrane as a ferroelectric gate dielectric, and showed that BTO membranes retain robust hysteretic behavior when interfaced with 2D TMDs. The large dielectric constant of BTO membranes effectively screen Coulomb scatterers, which leads to a large mobility \cite{Puebla2022}. While the primary focus of integrating oxide membranes has been using them as gate dielectric layers in FET devices to leverage their large dielectric constant, functionalities retained in oxide membranes such as ferroelectricity and their coupling to excitons in 2D TMDs at the heterogeneous 2D/3D interface remains poorly understood. 

In this work, we fabricate a dual gate device with monolayer MoSe$_2$ and a released BTO membrane, and study the influence of polarization on excitonic photoluminescence (PL) in monolayer MoSe$_2$. In the dual gate device, we achieve in-situ polarization switching in BTO and optical read-out of the polarization using excitons in interfacial MoSe$_2$. This eliminates the need of a separate pre-poling process and greatly simplifies the poling and read-out process. Polarization switching in BTO changes the relative population of different exciton species in MoSe$_2$. Our work is a step toward next-generation 2D/3D materials-based devices with tunable functionalities. Without the need for lattice matching, we achieve a heterogeneous 2D/3D interface where it is possible to electrically pole the device and observe corresponding ratio changes of neutral and charged excitons.

\section{\label{sec:level2} Results and Discussion}

\begin{figure*}
\includegraphics[width=\textwidth]{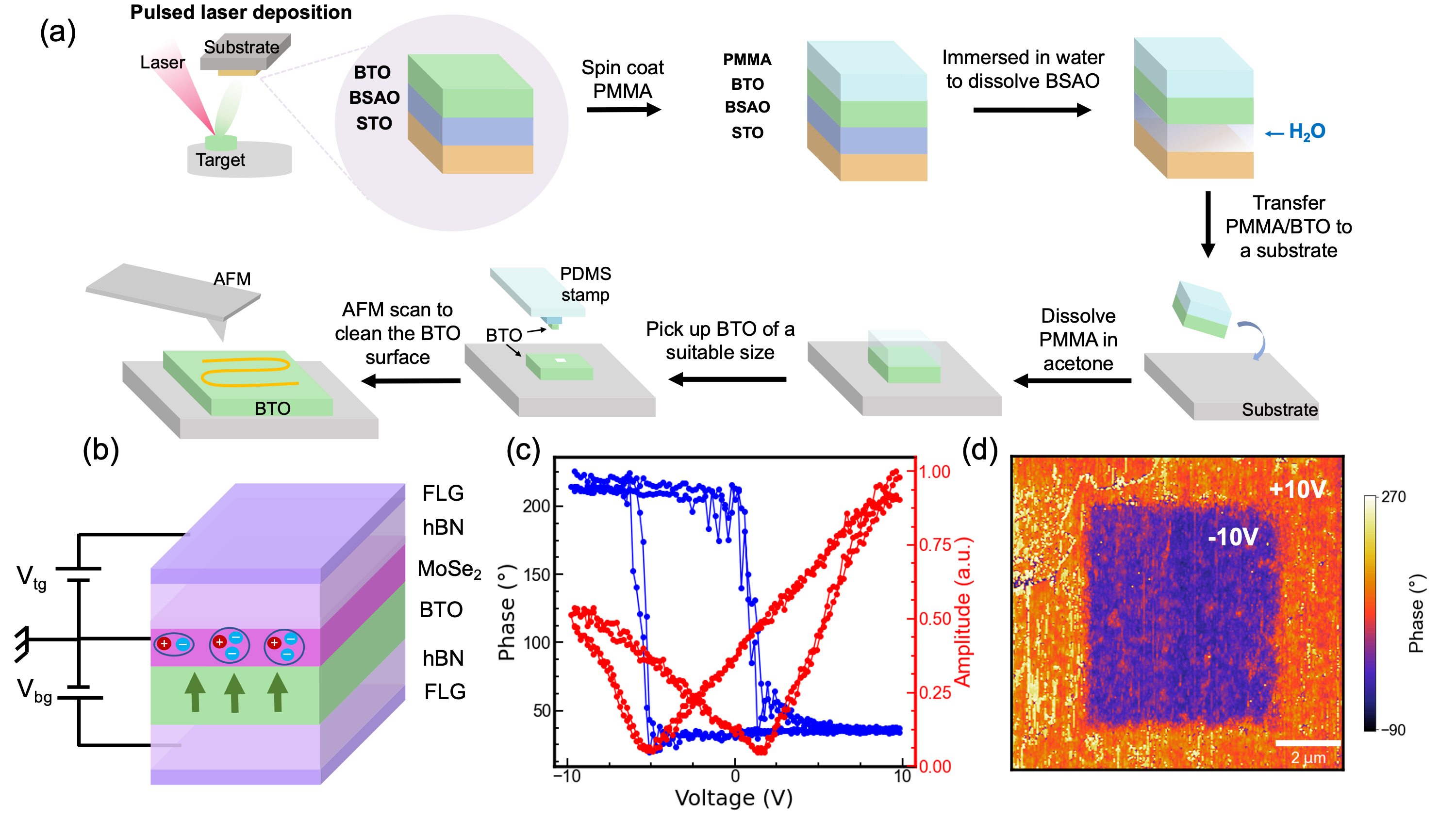}
\caption{\label{fig:wide} Device fabrication process and ferroelectric properties of a thin BTO membrane. (a) Schematic illustration of the release and the transfer of a BTO membrane and subsequent fabrication of a hybrid heterostructure. (b) Schematic diagram of a BTO/MoSe$_2$ based dual gate field-effect device. A 40~nm-thick BTO membrane is interfaced with monolayer MoSe$_2$. Top ($V_{tg}$) and bottom($V_{bg}$) gates are applied to form an out-of-plane electric field across the device. (c) Local phase hysteresis loop and butterfly amplitude curve acquired from a transferred BTO membrane (40~nm thick) stacked on a gold pad at 300~K. (d) PFM phase image after BTO (30~nm thick) is poled with a tip voltage of $\pm$10~V at 300~K.}
\end{figure*}

Figure 1a outlines the synthesis and transfer process of BTO membranes that have been lifted-off from their growth substrates. We grow 40~nm-thick BTO thin films that are grown epitaxially on single-crystal SrTiO$_3$ substrates with a 25~nm-thick Ba$_{0.44}$Sr$_{2.56}$Al$_2$O$_6$ (BSAO) water-soluble sacrificial layer via pulsed laser deposition. The BSAO layer can be dissolved in water and is used for BTO isolation from the growth substrate. In BSAO, Ba is substituted into water-soluble Sr$_3$Al$_2$O$_6$ for optimal lattice matching to the BTO films, thereby reducing strain in the BTO and improving membrane quality \cite{Singh2019_2}. Following the heterostructure synthesis, we spin coat a polymethyl methacrylate (PMMA) layer, forming a PMMA/BTO/BSAO/STO heterostructure and  immerse the heterostructure in DI water to etch the sacrificial layer. Once etching is complete, we transfer the PMMA/BTO structures onto a thermally oxidized Si substrate by using PMMA to make the membrane easier to handle. The substrate is then immersed in acetone to dissolve the PMMA, which leaves a large-area freestanding BTO membrane. Next we employ an additional pick-up process to obtain microscale area BTO membranes capable of dual gate device integration. For this we place a small piece of polydimethylsiloxane (PDMS) on the BTO membrane and peel it off of the underlying substrate to pick up a microscale piece of the BTO suitable for device integration. Following the pick-up, an AFM tip in contact mode scans the BTO surface to remove residual polymer, which is a key step to enhance the interface quality \cite{Goossens2012, Rosenberger2018, Kim2019, Hou2021, Chen2021} (See Methods in supporting information for more details). Figure 1b is a schematic diagram of a dual gate device. We apply top gate (V$_{tg}$) and bottom gate (V$_{bg}$) few-layer graphene (FLG) gates to create an electric field across the device. Top and bottom hexagonal boron nitride dielectrics are symmetric in thickness, and we apply V$_{tg}$ and V$_{bg}$ of the same magnitude but the opposite polarity to switch the polarization in BTO. Also, we pulse the voltage to study the remnant polarization, where V$_{poling}$ $=$ V$_{top}$ $=$ $-$V$_{bottom}$.

We conduct piezoresponse force microscopy (PFM) hysteresis measurements to confirm the ferroelectricity in the released BTO membrane. Local phase hysteresis and amplitude butterfly-shaped curves in figure 1c confirm that BTO retains its out-of-plane (OOP) ferroelectricity even after its removal from the oxide substrate. We find that the coercive voltages vary from region-to-region (Fig.~S3).  We speculate that the variable hysteresis observed in BTO may be attributed to irregularities in strain that arise from the uneven surface of gold pads, wrinkles, and polymer residue \cite{Choi2004, Kim2022}. The presence of impurities and polymer residue lead to fluctuations in the tip-surface interaction during the poling process. Additionally, we write domains via PFM tip on a BTO membrane by applying an opposite tip voltage ($\pm$10~V). The phase contrast image in figure 1e shows 180$^{\circ}$ polarization switching in the squared region. At room temperature, BTO has tetragonal ferroelectric structure with polarization pointed along the long axis. Our membranes have this polarization oriented OOP due to the compressive strain applied by the STO substrate during synthesis. Even at temperatures below 20~K, BTO is known to preserve its OOP polarization, with a transition to a ferroelectric rhombohedral structure below 180~K \cite{Shu2001}. 

\begin{figure*}
\includegraphics[width=\textwidth]{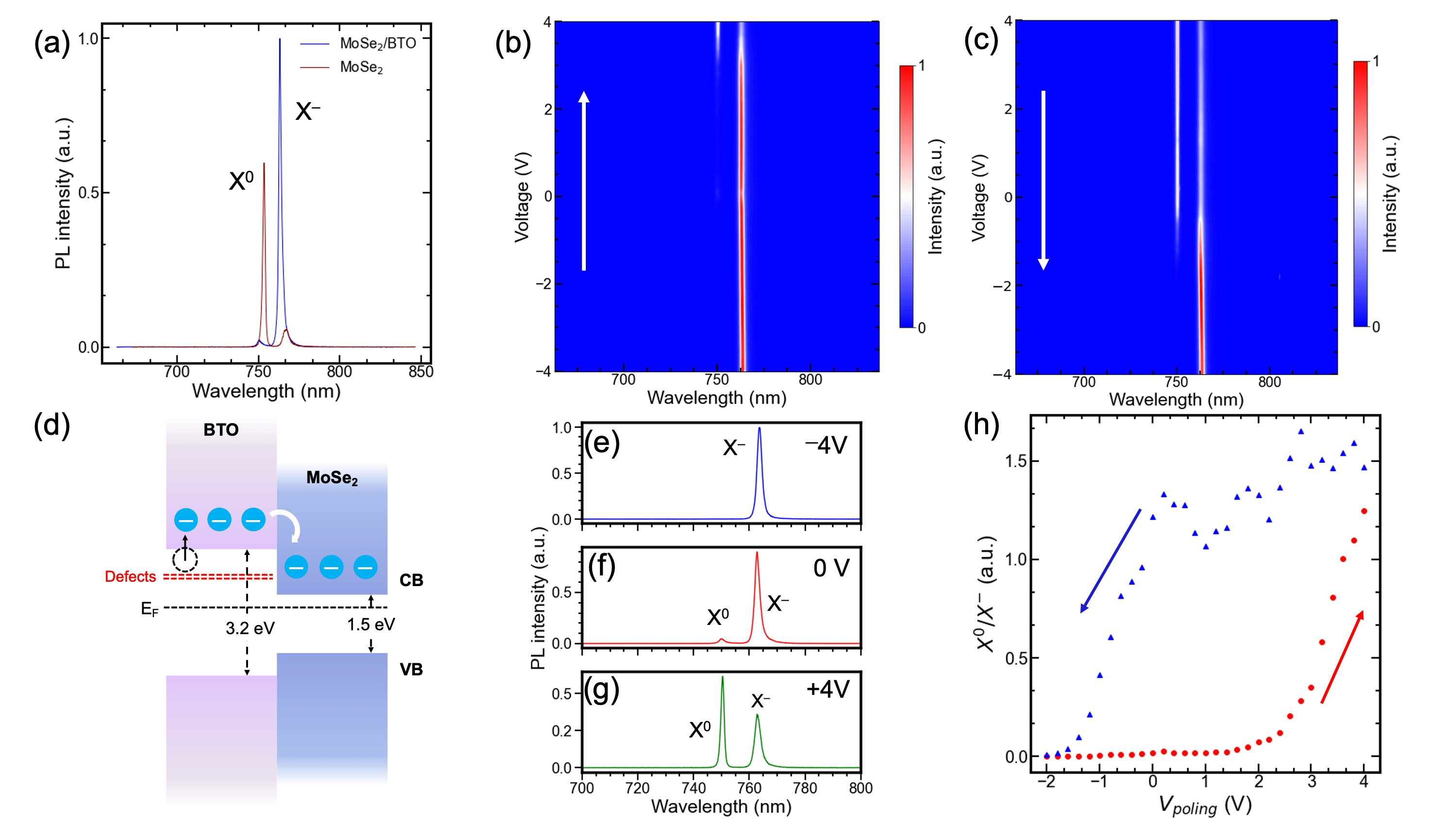}
\caption{\label{fig:wide}Ferroelectricity-induced PL characteristics in MoSe$_2$ at 15~K (a) PL spectra from a 1L-MoSe$_2$/BTO (blue) heterostructure and hBN capped 1L-MoSe$_2$ alone (red) at 15~K. (b) PL variation of MoSe$_2$ at 15~K when scanning the pulsed voltage from $-$4~V to $+$4~V. (c) As the voltage sweep direction is reversed, the PL shows hysteretic behavior. (d) Energy band diagram of MoSe$_2$/BTO, showing defect levels (including Ti$^{3+}$ ions and oxygen vacancies) located just below the conduction minimum in BTO, which act as a donor level. This drives the interfacial charge transfer from the BTO conduction band to the MoSe$_2$ conduction band. (e)-(g) Spectra at $V_{tg}$ (−$V_{bg}$) equal to $-$4~V, 0~V, +4~V, respectively. (h) Poling voltage dependent X$^0/$X$^{−}$ shows hysteretic behavior.}
\end{figure*}

Next, we measure the photoluminescence (PL) of the device. We pump the MoSe$_2$ layer with a 532~nm laser with 1.8~$\mu$W power at 15~K. Figure 2a shows the PL spectra of hBN capped monolayer MoSe$_2$ (1L-MoSe$_2$)/BTO heterostructure and 1L-MoSe$_2$ alone. The PL spectrum of bare 1L-MoSe$_2$ shows dominant X$^0$ emission and weak X$^{-}$ emission, which implies that MoSe$_2$ is slightly n-doped due to impurities attached to reactive chalcogenide vacancies or defects in the substrate \cite{Komsa2012, Singh2019, Duan2021}. The emission from 1L-MoSe$_2$/BTO heterostructure is slightly blue-shifted, and it shows enhanced X$^{-}$ emission, which is driven by n-type doping in BTO. BTO typically shows n-type behavior due to oxygen vacancy induced Ti$^{3+}$ ions \cite{Chandrappa2023}. Ti$^{3+}$ ions form a shallow donor level located close to the conduction band minimum together with other oxygen defects such as oxygen vacancies most common in perovskite oxides \cite{Sagdeo2018, Smyth1985}, and this increases the electron concentration in the BTO conduction band \cite{Amano2016, Valentin2009}. The PL does not show an additional emission peak which is attributable to interlayer excitons (IXs). Moreover, the trion intensity enhancement observed when MoSe$_2$ is in contact with BTO, combined with the absence of IXs in the spectrum, suggests the type I band alignment shown in Figure 2d \cite{Ying2020, Zhang2021}. BTO has a wider band gap ($\sim$3.2 eV) than that of MoSe$_2$ ($\sim$1.5 eV), and when this BTO is in contact with MoSe$_2$, excess electrons in the BTO conduction band relax into the MoSe$_2$ conduction band. The large electron population in MoSe$_2$ facilitates the binding of electrons with neutral excitons (X$^{0}$) to form negatively charged trions (X$^{-}$), leading to X$^{-}$ emission enhancement. 

\begin{figure*}
\includegraphics[width=\textwidth]{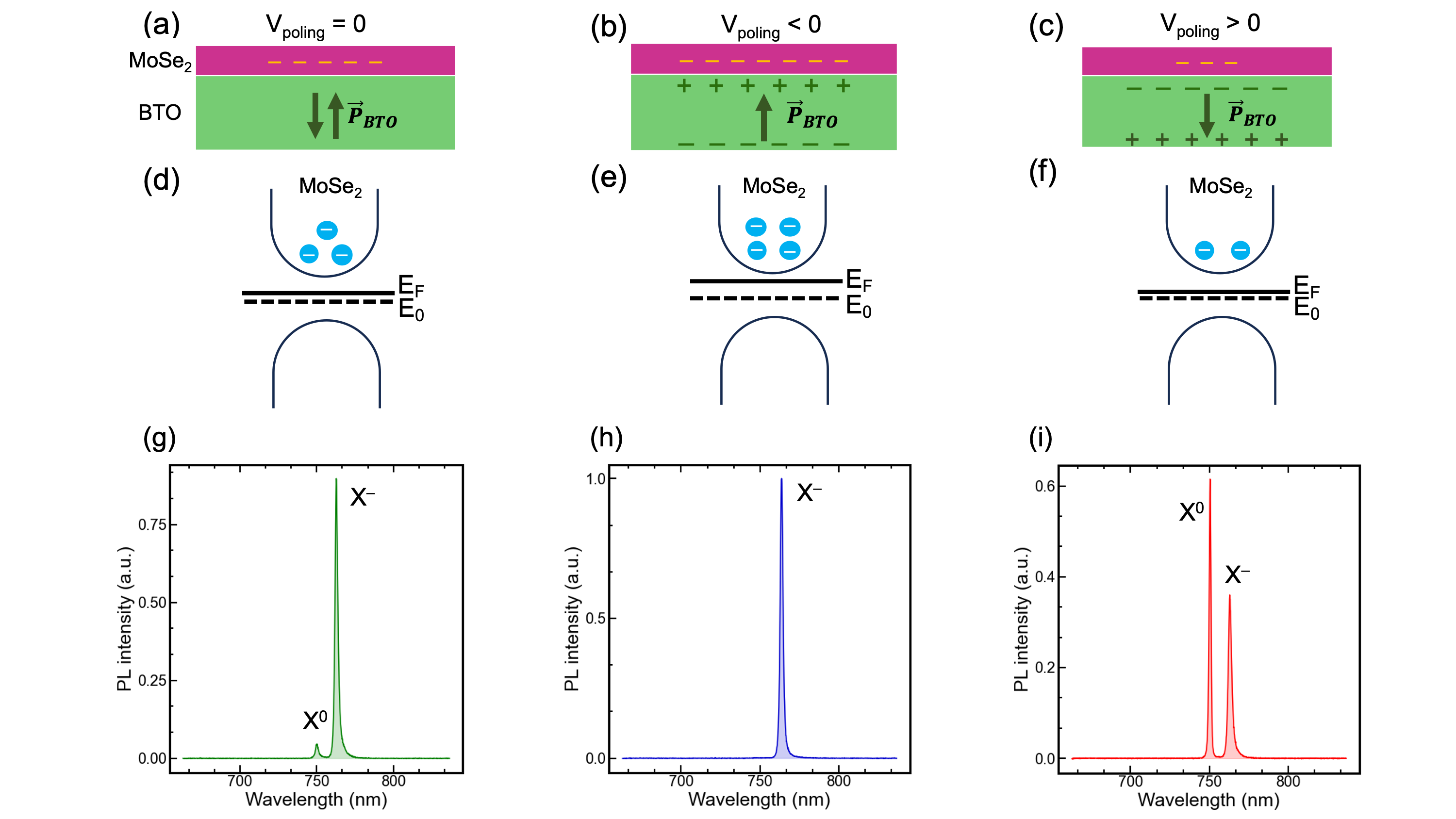}
\caption{\label{fig:wide} Diagram showing how polarization switching alters charge accumulation and PL. (a-c) $V_{poling}$ induced change in BTO polarization. (d) Initially, the polarization is in random state, and MoSe$_2$ is slightly n-type due to the charge transfer from BTO. (e) Upward polarization raises the Fermi level closer to the conduction band minimum (n-doping) (f) Downward polarization lowers the Fermi level toward the valence band maximum (p-doping) (g-i) The PL spectra each reflect the corresponding polarization shown in (d-f).}
\end{figure*}

Next, we study the remnant polarization induced X$^0$ and X$^{-}$ emission in MoSe$_2$. We apply a pulsed voltage to switch the polarization in BTO (Fig.~S4) and then acquire a PL spectrum at 15~K. Figure 2b shows a PL change when we scan the pulsed voltage from $-$4~V to $+$4~V. At first, X$^{-}$ emission is dominant because MoSe$_2$ is heavily n-doped when it is integrated with BTO. When $+$3~V pulse voltage is applied, X$^{0}$ emission starts to be strengthened, which implies that more holes are induced in MoSe$_2$ that bind with free electrons to form X$^{0}$. This makes MoSe$_2$ slightly less n-doped. The voltage sweep direction is reversed in figure 2c, and the PL change shows a different behavior. While the enhancement of X$^{0}$ happens at V$=+$3 when varying the voltage from $-$4~V to $+$4~V, the switch to a X$^{0}$ dominated spectrum occurs at V$=-$0.8, when reversing the voltage sweep direction.  This asymmetry in PL variation is a fingerprint of the ferroelectric hysteresis. 

The PL spectra at different poling voltages shown in figure 2e-g demonstrates the relative X$^{0}$ and X$^{-}$ intensity change more clearly. When V$_{tg}=$ 0 (Fig.~2f), we observe weak X$^{0}$ and dominant X$^{-}$ emission peaks, showing the n-type doping in MoSe$_2$. When V$_{tg}<$ 0 is applied, X$^{-}$ emission is further enhanced and X$^{0}$ emission peak is suppressed (Fig.~2e), indicating that MoSe$_2$ is heavily n-doped. When V$_{tg}>$ 0 is applied, X$^{0}$ emission intensity increases, which implies that MoSe$_2$ is less n-doped (Fig.~2g). We conduct the same pulsed measurement on the dual-gate device devoid of BTO (Fig.~S7). In contrast to the dual-gate device with BTO, the control device exhibits no spectral change upon pulsing and reversal of the pulse voltage polarity. This lack of hysteresis further verifies that the observed change in PL characteristics stems from the ferroelectric properties inherent in BTO, ruling out potential influences from factors like contact resistance. 

The ratio between X$^0$ and X$^{-}$ can indicate polarization switching-induced doping in MoSe$_2$. If the ratio variation originates from the polarization switching in BTO, we would expect to see hysteretic behavior in the relative change of X$^0$ and X$^{-}$ emission intensity as we vary the poling voltage. To ascertain the voltage dependent X$^{0}/$X$^{-}$ behavior, we extract the emission intensities of X$^0$ and X$^{-}$ by fitting the excitonic lines to Gaussians, and plot the area ratio of  X$^0$ and X$^{-}$ against the poling voltage (Fig.~2h). As seen in figure 2h, X$^0/$X$^{-}$ shows hysteresis. This unambiguously shows that the ferroelectricity in BTO is responsible for the charge accumulation in MoSe$_2$ and that excitons capture polarization-induced changes in doping. Moreover, this read-out is viable even at room temperature (refer to more details in supporting information).

The mechanism of polarization-induced doping and the resulting PL variation is illustrated in figure 3a-i. Before applying a poling voltage, there is a mixture of up and down polarization states in BTO (Fig.~3a). Even before the poling, MoSe$_2$ is slightly n-doped due to the charge transfer from BTO and the Fermi level (E$_F$) is located close to the conduction band minimum (Fig.~3d). This is reflected in prominent X$^{-}$ emission in the spectrum (Fig.~3g). Upon application of a negative poling voltage, the polarization in BTO is switched upward (Fig.~3b). The upward polarization induces holes to accumulate at the top surface of BTO. This attracts more electrons to screen out the hole population near the MoSe$_2$ and BTO interface, elevating E$_F$ of MoSe$_2$ closer to the conduction band minimum (Fig.~3h). The increase in free electrons facilitates electrons binding with X$^0$ to form X$^-$, thereby amplifying the X$^-$ emission intensity (Fig.~3h). Conversely, a positive poling voltage switches down the polarization (Fig.~3c), and this pushes E$_F$ closer to the valence band maximum (Fig.~3f). This reduces the free electron population in MoSe$_2$ conduction band and fosters more X$^0$ formation, leading to X$^0$ enhancement in PL (Fig.~3i). It is important to note that MoSe$_2$ is heavily n-doped with electrons transferred from BTO upon interfacing MoSe$_2$ and BTO. Converting its doping type from n to p would require a significantly higher gate voltage than it is needed to switch the polarization in BTO. 

\begin{figure*}
\includegraphics[width=\textwidth]{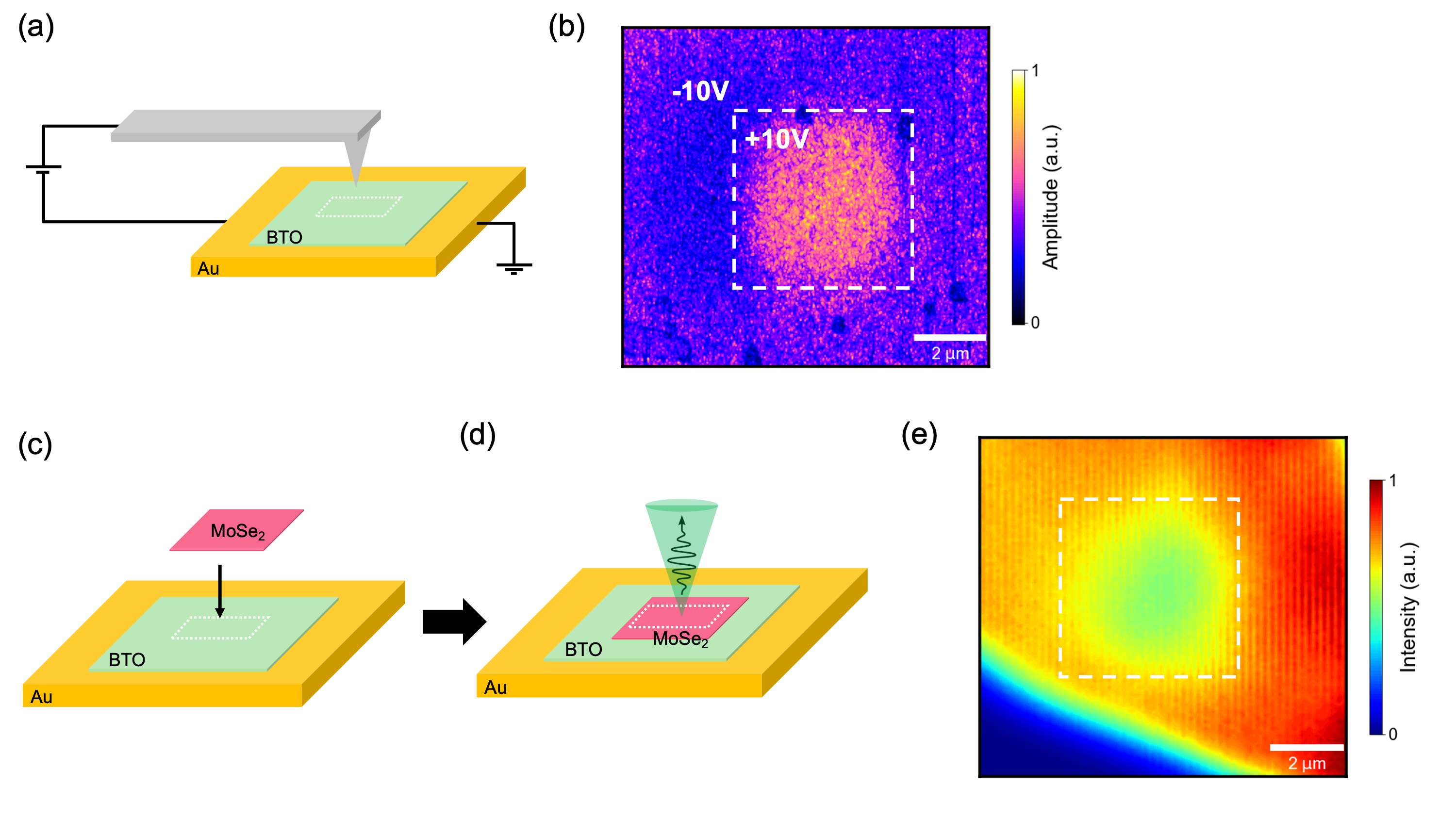}
\caption{\label{fig:wide} PFM domain writing of BTO and corresponding PL characteristics at room temperature. (a) Schematic diagram of the domain writing setup. We stack BTO on top of a gold electrode and apply a tip voltage in a squared region (white dotted line) to switch the polarization. (b) PFM amplitude image after writing ferroelectric domains on BTO. (c) Monolayer MoSe$_2$ is stacked on top of BTO immediately after the domain writing (d) Subsequent PL measurement on MoSe$_2$ after its transfer. (e) PL variation in MoSe$_2$ stacked on the pre-poled BTO membrane at room temperature. }
\end{figure*}

To conclusively confirm that the primary reason for the PL variation is the polarization in BTO, we apply a local electric field with a PFM tip to write ferroelectric domains in BTO. Subsequently, we examine how the MoSe$_2$ PL characteristics change after transferring MoSe$_2$ onto the domain-engineered BTO. Figure 4a illustrates a schematic representation of the PFM domain writing setup. Using the PFM tip, we perform box-in-box domain writing on a 40~nm thick BTO membrane situated on a gold pad at room temperature. The PFM amplitude image captured shortly after the domain writing is shown in figure~4b. A distinctive amplitude contrast between inner and outer boxes indicate antiparallel polarization flipping within each box. Following the domain writing, we immediately transfer MoSe$_2$ on top of the domain engineered BTO. It is important to note that the sample temperature is maintained below 60 $^{\circ}$C when we do the transfer, well below the critical temperature of BTO. After the transfer, we examine its PL at room temperature (Fig.~4c). As shown in figure 4d, a noticeable PL change in MoSe$_2$ after its placement on the domain engineered BTO membrane. The decrease in PL intensity within the squared region outlined by a dotted line is attributed to the downward polarization in BTO directed towards the gold substrate. At room temperature, we observe a reduction in the integrated PL intensity concurrent with an increase in trion population. This is because trions take non-radiative decay path, leading to diminished PL intensity. This is matched with the room temperature PL variation (Fig. S9). Additionally, the intensity contrast evident in the PL map in figure 4d aligns with the amplitude image in figure 4b, confirming that the PL spectral change indeed arises from a change to the polarization of BTO. This verification also underscores that our dual-gate device facilitates in-situ polarization switching and simultaneous optical read-out of the induced doping level, eliminating the necessity for a separate pre-poling process and greatly streamlining the experiment. 

\section{Conclusion}
\raggedbottom
In summary, we have integrated a thin isolated BTO membrane with monolayer MoSe$_2$ in a dual gate device. BTO membranes that have been lifted off from their growth substrates allow for streamlined integration with 2D materials, enabling a 2D/3D heterogeneous interface in a wide array of structures. Furthermore, BTO membranes retain their OOP ferroelectricity even after being removed from their substrates. The coupling of BTO polarization with monolayer MoSe$_2$ enables modification of electronic and PL properties in MoSe$_2$. In the dual gate device, flipping the polarization can be achieved using device electrodes, and the resulting change in polarization-induced doping level is concomitantly read-out by the X$^0$ and X$^-$ ratio change. The hysteresis observed in X$^0$/X$^-$ and the PL variation at a heterogenous MoSe$_2$/BTO interface confirms that the BTO and MoSe$_2$ are able to form a clean heterointerface that allows charge transfer and coupling of the polarization and excitons. Our results show a substantial step toward the device-level application of 2D TMD/3D perovskite oxide heterostructures for compact and flexible devices.

\section{Supporting Information}
Detailed description of experimental methods and additional figures (PDF).

\section{Acknowledgements}
This work is primarily supported by AFOSR MURI (FA9550-18-1-0480). We also acknowledge funding from the Department of Energy, Office of Basic Energy Sciences, and Division of Materials Sciences and Engineering (DE-AC02-76SF00515). K.W. and T.T. acknowledge support from the JSPS KAKENHI (Grant Numbers 21H05233 and 23H02052) and World Premier International Research Center Initiative (WPI), MEXT, Japan.

\nocite{*}

\bibliography{Jaehong}

\end{document}


\preprint{APS/123-QED}

\title{Supporting Information: \\
Tuning exciton emission via ferroelectic polarization at a heterogeneous interface between a monolayer transition metal dichalcogenide and a perovskite oxide membrane}

\author{Jaehong Choi}
\email{jc3452@cornell.edu}
\affiliation{School of Applied and Engineering Physics, Cornell University, Ithaca, NY 14850, USA}

\author{Kevin J. Crust}%
\affiliation{Department of Physics, Stanford University, Stanford, CA 94305, USA}
 \affiliation{Stanford Institute for Materials and Energy Sciences,
SLAC National Accelerator Laboratory, Menlo Park, CA 94025, USA}
 
\author{Lizhong Li}
\affiliation{School of Applied and Engineering Physics, Cornell University, Ithaca, NY 14850, USA}

\author{Kihong Lee}
\affiliation{Department of Physics, Cornell University, Ithaca, NY 14850, USA}

\author{Jialun Luo}
\affiliation{Department of Physics, Cornell University, Ithaca, NY 14850, USA}

\author{Jae Pil So}
\affiliation{School of Applied and Engineering Physics, Cornell University, Ithaca, NY 14850, USA}

\author{Kenji Watanabe}
\affiliation{Research Center for Electronic and Optical Materials, National Institute for Materials Science, 1-1 Namiki, Tsukuba 305-0044, Japan}

\author{Takashi Taniguchi}
\affiliation{Research Center for Electronic and Optical Materials, National Institute for Materials Science, 1-1 Namiki, Tsukuba 305-0044, Japan}

\author{Harold Y. Hwang}
\affiliation{Stanford Institute for Materials and Energy Sciences,
SLAC National Accelerator Laboratory, Menlo Park, CA 94025, USA}
 \affiliation{Department of Applied Physics, Stanford University, Stanford, CA 94305, USA}

\author{Kin Fai Mak}
\affiliation{School of Applied and Engineering Physics, Cornell University, Ithaca, NY 14850, USA}
 \affiliation{Department of Physics, Cornell University, Ithaca, NY 14850, USA}
  \affiliation{Kavli Institute at Cornell for Nanoscale Science, Ithaca, NY 14850, USA}

\author{Jie Shan}
\affiliation{School of Applied and Engineering Physics, Cornell University, Ithaca, NY 14850, USA}
 \affiliation{Department of Physics, Cornell University, Ithaca, NY 14850, USA}
  \affiliation{Kavli Institute at Cornell for Nanoscale Science, Ithaca, NY 14850, USA}

\author{Gregory D. Fuchs}
\email{gdf9@cornell.edu}
\affiliation{School of Applied and Engineering Physics, Cornell University, Ithaca, NY 14850, USA}
\affiliation{Kavli Institute at Cornell for Nanoscale Science, Ithaca, NY 14850, USA}

\newpage

\maketitle


\section{Methods}
\subsection{BTO thin film synthesis and isolation}
We synthesize a 40~nm-thick BaTiO$_3$(BTO)$/$Ba$_{0.44}$Sr$_{2.56}$Al$_2$O$_6$ heterostructures on (001)-oriented SrTiO$_3$ substrates using the pulsed laser deposition. The Ba$_{0.44}$Sr$_{2.56}$Al$_2$O$_6$ layer was first grown with a substrate temperature of 730$^{\circ}$C, an argon pressure of $4 \times 10^{-6}$~Torr, a laser fluence of 2.0~J~cm$^{-2}$, a laser repetition rate of 1~Hz, and an imaged laser spot size of 4.2~mm$^{2}$. The BTO layer was then grown with a substrate temperature of 680$^{\circ}$C, an oxygen pressure of 20~mTorr, a laser fluence of 2.75~J~cm$^{-2}$, a laser repetition rate of 1~Hz, and an imaged laser spot size of 2.1~mm$^{2}$, with the sample being cooled to room temperature at the growth pressure after synthesis. We then spin coat a PMMA layer on top of the heterostructure, cut the sample into smaller pieces (3~mm $\times$ 3~mm size), and immerse the heterostructure in water for approximately 48 hours to etch away the sacrificial layer. After the etching, we lift the PMMA/BTO structure to the surface of the water using tweezers and pick it up using a water scoop. The PMMA/BTO structure is transferred onto a thermally oxidized silicon (285~nm SiO$_2$) substrate and heated to 60 $^{\circ}$C to remove excess water, with the sample finally being placed in acetone to dissolve the PMMA from the surface of the membrane. 

\subsection{BTO membrane integration in a dual-gate device}
We use a PDMS (PF-3-X4, Gel-Pak) stamp to pick up small BTO membranes (approximately 20 $\mu$m $\times$ 20 $\mu$m area) for the device integration. To pick up the small BTO membranes, we place a PDMS film on the large BTO membrane and gently peel it off. After the pick-up, the PDMS stamp is investigated under an optical microscope. Once we identify the membranes with the suitable size, we transfer them onto the target substrate where hBN/bottom FLG layers are pre-stacked, by using an all-dry transfer method~\cite{Gomez2014}.

\subsection{AFM cleaning of BTO membranes}
To improve the BTO/MoSe$_2$ interface for their better coupling, we scan the BTO surface slowly with an AFM tip (AC200TS, Oxford Instruments) in contact mode before integrating with MoSe$_2$ over a 12-hour period. After the AFM cleaning, we stack MoSe$_2$/hBN/top FLG onto the AFM cleaned BTO by using a layer-by-layer transfer method~\cite{Wang2013}. 

\subsection{Piezoresponse force microscopy measurement}
We perform PFM measurements using the Asylum-MFP3D-Bio-AFM-SPM under ambient conditions. In the PFM measurement setup, BTO is pre-stacked on a gold pad patterned on a Si substrate. We use the MFP3D Lithography PFM mode for domain writing. A bias voltage is applied to Pt coated (30~nm) commercial tips (HQ:NSC18/Pt) to inscribe a box-in-box patterned ferroelectric domain on BTO. Subsequent to the writing, amplitude and phase information are collected. For the local hysteresis measurement, a tip voltage with a triangle step function waveform is applied through the tip. The phase and amplitude changes of BTO are then measured each time the tip voltage amplitude goes down to zero. 

\subsection{Photoluminescence measurements}
We perform PL measurements with our home-built laser-scanning confocal microscope including a Princeton instruments spectrometer (PyLon 100BR). We pump the device with a 532~nm laser (Dragon lasers) at a power below 5~$\mu$W while we cool down the device inside a Janis He-flow cryostat, which is capable of maintaining the sample temperature ranging from 15~K to 300~K. 

\subsection{Pulsed voltage measurement}
We use a Keithley 2400 and 2401 to apply pulsed voltages across gold electrodes fabricated on a SiO$_2$ (285~nm)/Si substrate (WaferPro). We fabricate these electrodes by sequentially depositing 5~nm of Cr (adhesion layer) and 50~nm of Au using an ebeam-evaporator. The pulse voltage duration is 5 seconds, and we conduct the spectral measurements 2 seconds after ramping the voltage down to zero. 

\setcounter{figure}{0}
\renewcommand{\figurename}{Fig.}
\renewcommand{\thefigure}{S\arabic{figure}}

\section{Supplemental figures}

\begin{figure*}[h!]
\includegraphics[width=\textwidth]{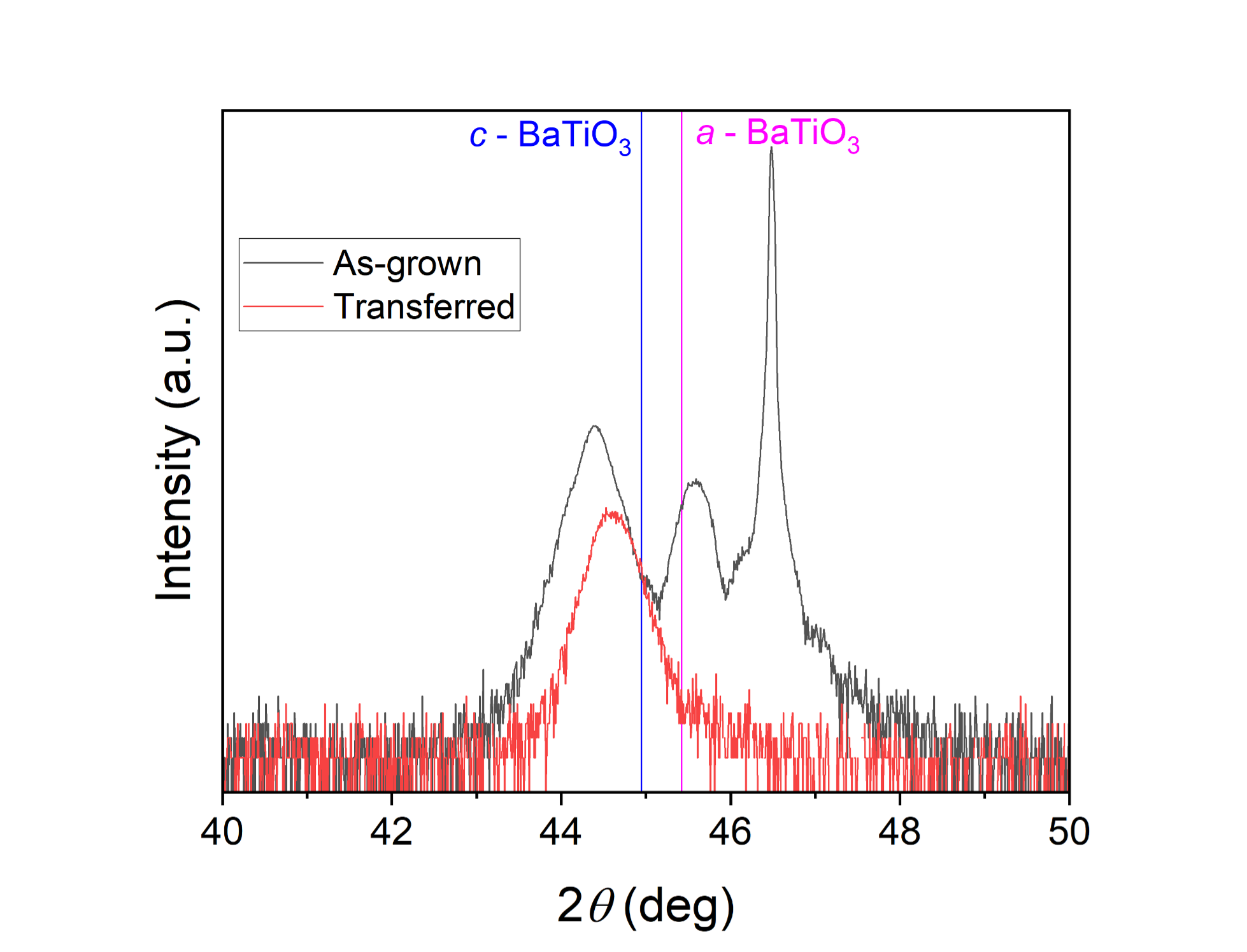}
\caption{\label{fig:wide} $\theta - 2\theta$ X-Ray Diffraction of BaTiO$_3$. The as-grown films are synthesized with the long $c-$axis out-of-plane due to compressive strain from the SrTiO$_3$ substrate. A slight lattice expansion can be observed, likely due to a combination of the Poisson effect and defects. The Ba$_{0.44}$Sr$_{2.56}$Al$_2$O$_6$ peak is observed to be quite close to the bulk $a-$axis of BaTiO$_3$ as intended, since it should be lattice matching to our in-plane BaTiO$_3$ lattice parameters. After release and transfer, the freestanding BaTiO$_3$ membrane maintains its $c-$axis orientation with a small lattice relaxation due to the removal of epitaxial strain.} 
\end{figure*}
\newpage

\afterpage{%
\begin{figure*}
\includegraphics[width=\textwidth]{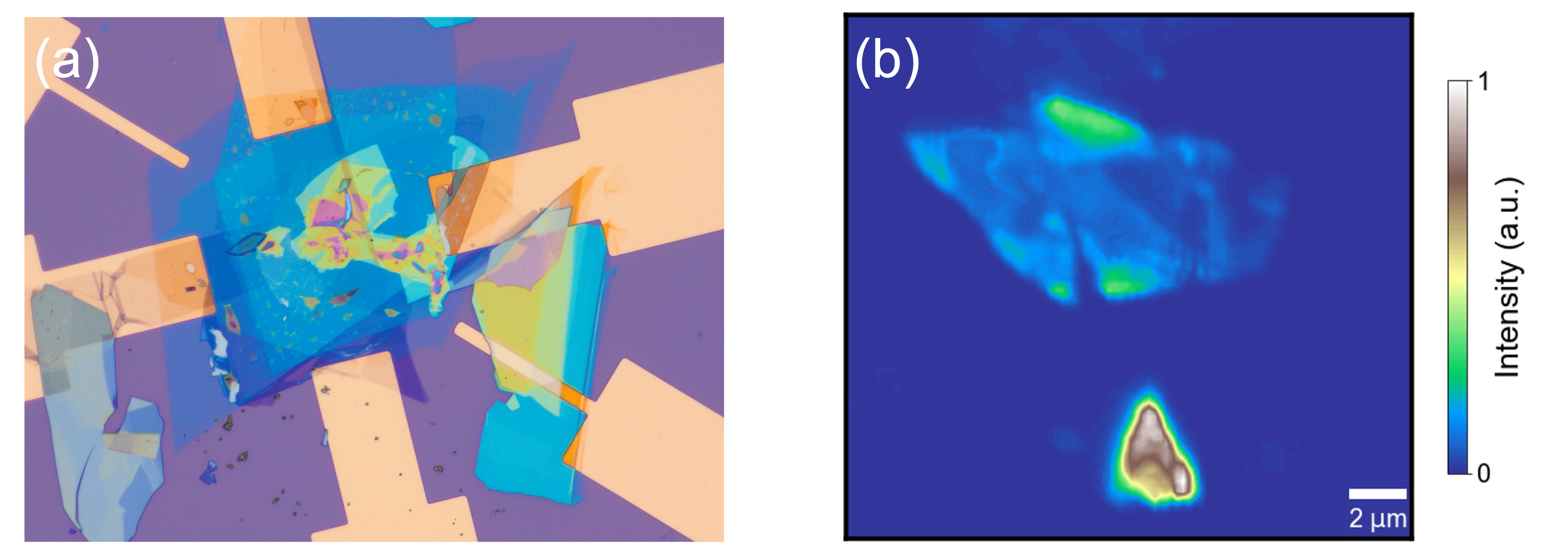}
\caption{\label{fig:wide} Optical image and PL map of the heterostructure. (a) Optical image of the device at room temperature. (b) PL image of the device at 15K.} 
\end{figure*}
\clearpage
}

\afterpage{%
\begin{figure*}
\includegraphics[width=\textwidth]{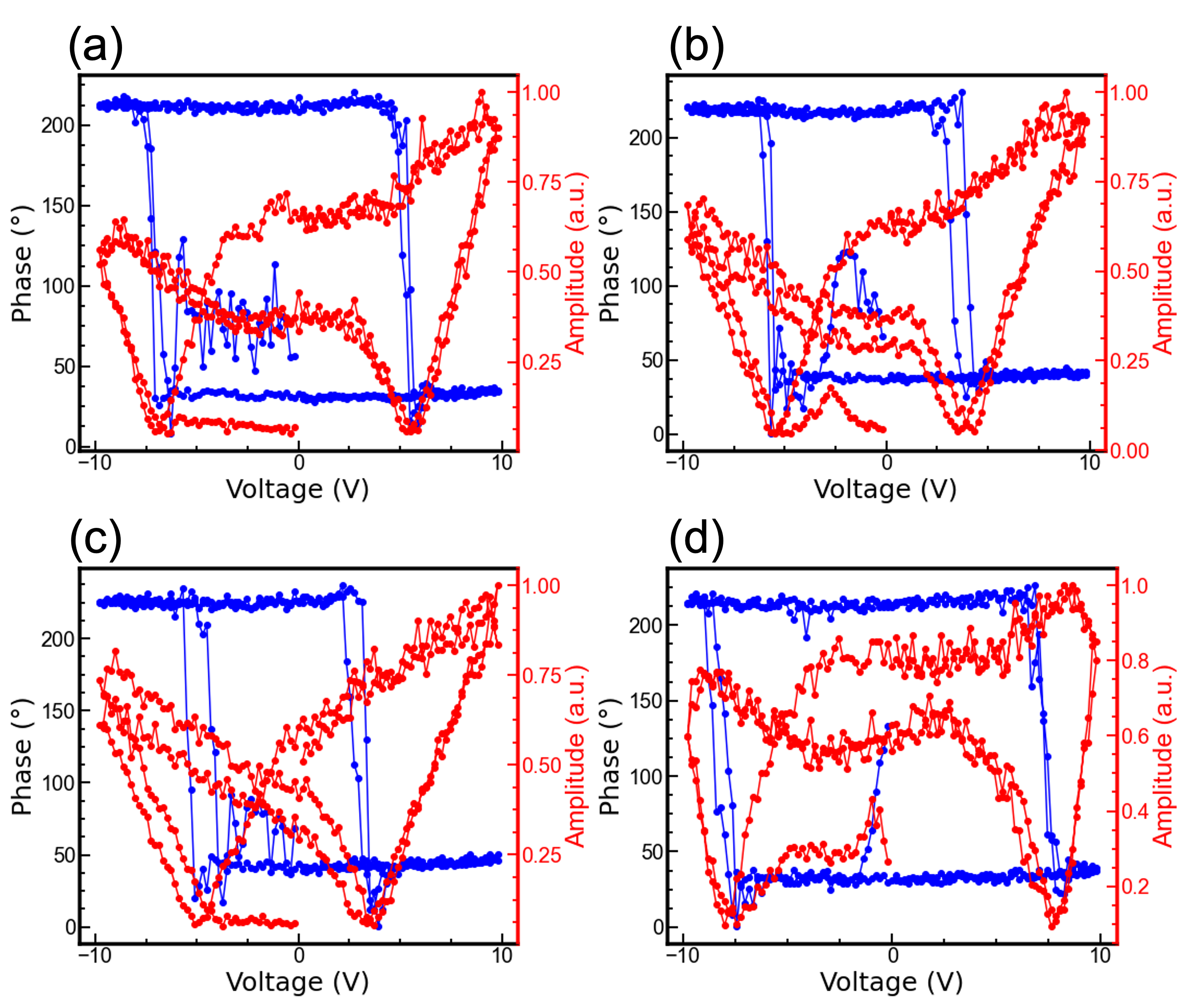}
\caption{\label{fig:wide} Phase and amplitude hysteresis loops at room temperature. (a),(b),(c),(d) Phase and hysteresis loops from the same BTO membrane show different coercive voltages, which is due to non-uniform lattice strain distributed across the membrane~\cite{Choi2004, Kim2022}.} 
\end{figure*}
\clearpage
}

\afterpage{%
\begin{figure*}
\includegraphics[width=\textwidth]{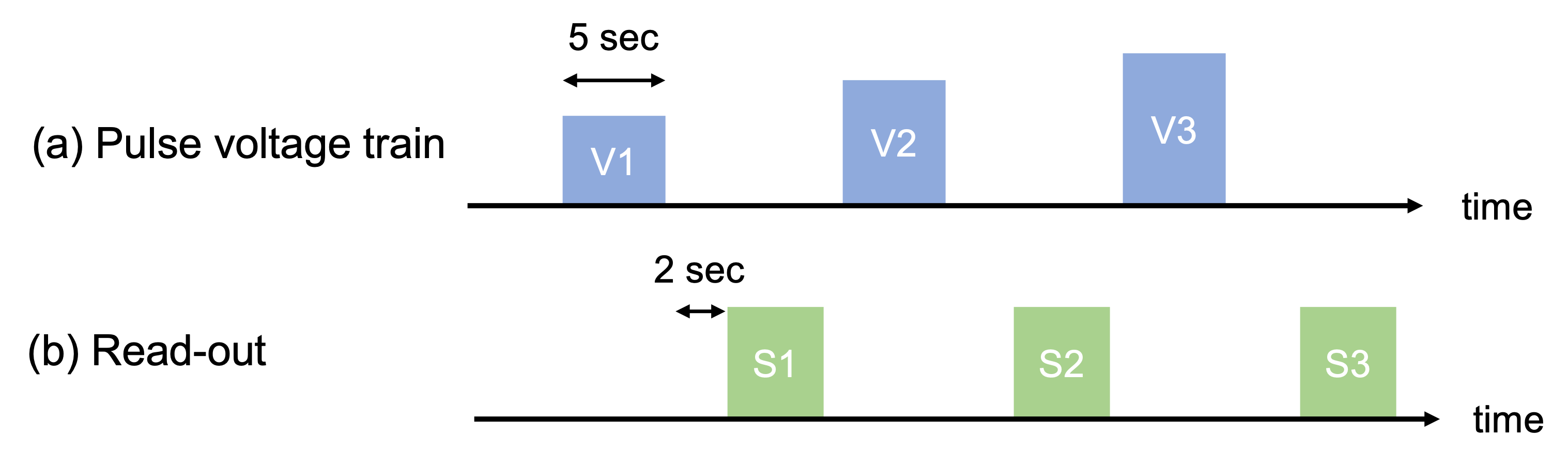}
\caption{\label{fig:wide} Pulse mode schematics. (a) We apply a pulse voltage train with each pulse duration of 5 seconds to the device. It is important to note that the actual ramp rate of the voltage is not as steep as depicted in the figure. (b) We measure PL spectrum 2 seconds after each pulse voltage ramps down to zero.} 
\end{figure*}
\clearpage
}

\afterpage{%
\begin{figure*}
\includegraphics[width=\textwidth]{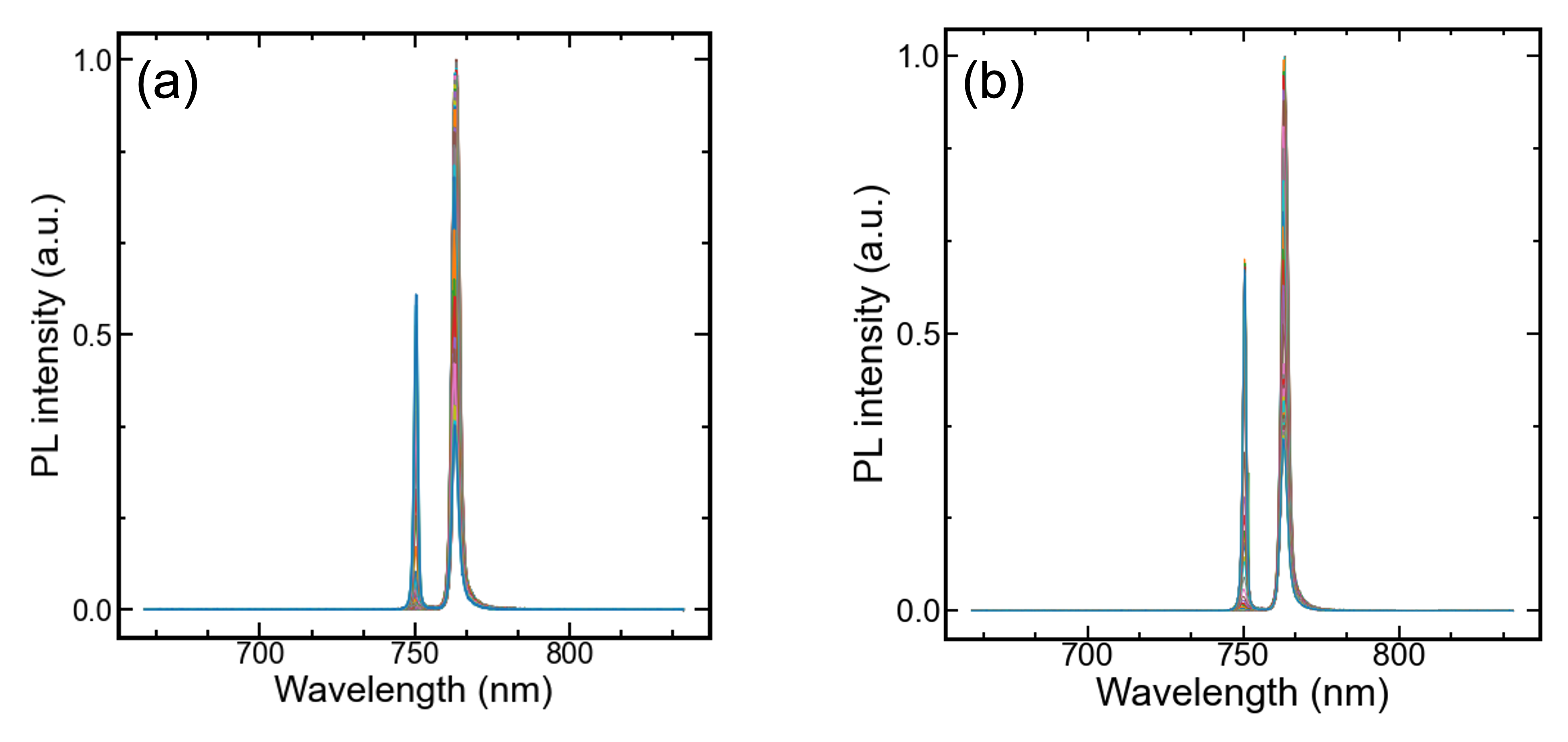}
\caption{\label{fig:wide} Spectral change when scanning the pulse voltage at 15~K. Spectra change (a) when the voltage is scanned from $–4$~V to $+4$~V, (b) when the voltage is scanned from $+4$~V to $–4$~V.} 
\end{figure*}
\clearpage
}

\afterpage{%
\begin{figure*}
\includegraphics[width=\textwidth]{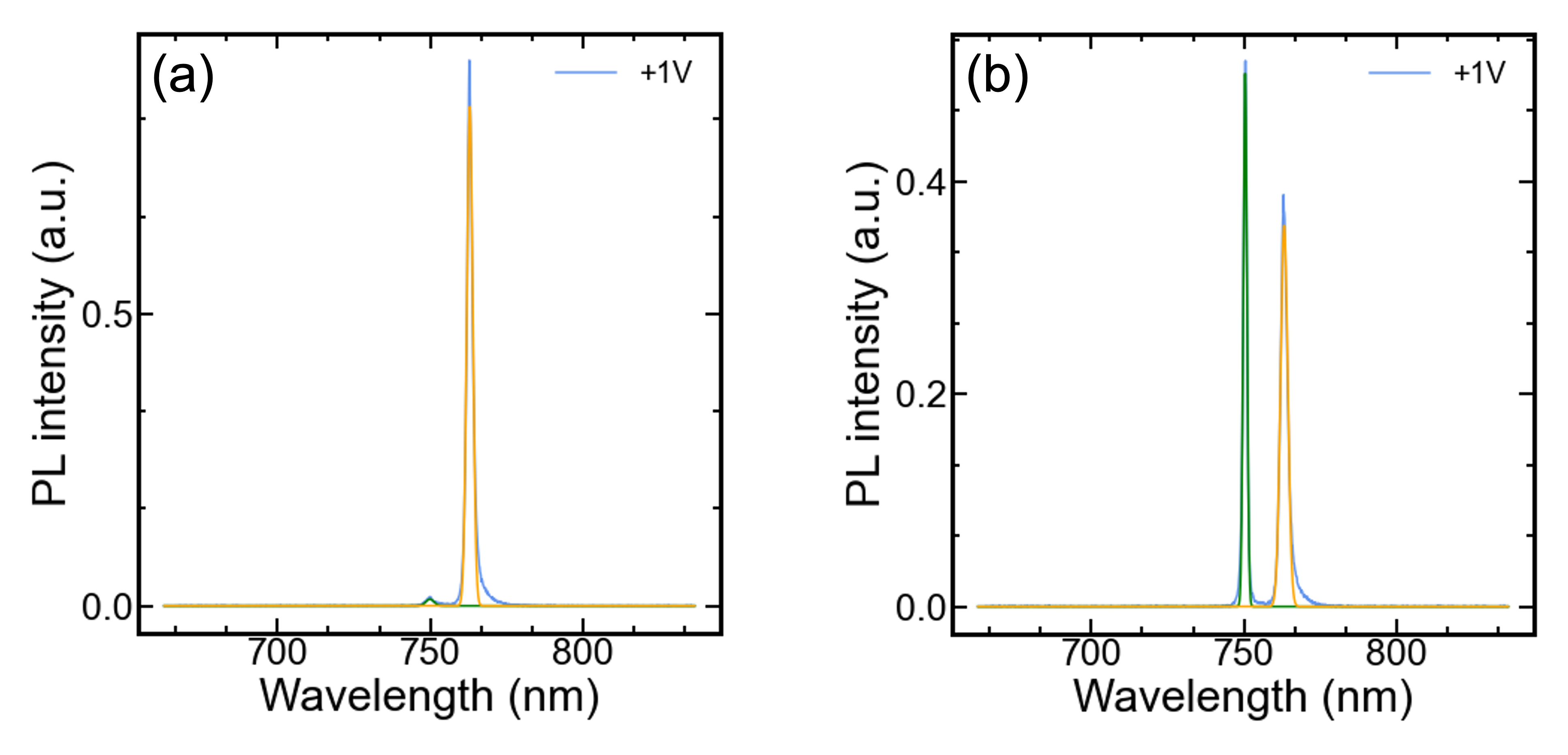}
\caption{\label{fig:wide} Gaussian fitting the area under spectra at 15~K. (a) spectrum when we apply V$_{poling}$ = $+1$~V while scanning V$_{poling}$ from $–4$~V to $+4$~V and (b) from  $+4$~V to $–4$~V.} 
\end{figure*}

Law of mass action explains the exciton concentration dependence of the charge carrier density ($n_e$) in semiconductors~\cite{Siviniant1999, Peimyoo2014,Duan2021}:

\begin{equation}
    \frac{N_{X^{-}}}{N_{X^{0}}} = \frac{\pi \hbar^2 m_{X^{-}} }{4 m_{X^0} m_e} \frac{e^{\frac{E_B}{k_{B} T}}}{k_{B} T} n_{e}
\end{equation}

\noindent where $N_{X^{0}}$ and $N_{X^{-}}$ are charge concentration of X$^{0}$ and X$^{-}$ respectively, $m_{X^{0}}$, $m_{X^{-}}$, and $m_e$ respectively are the effective mass of X$^{0}$, X$^{-}$, and electron, $\hbar$ is the reduced Planck's constant, $k_B$ is the Boltzmann constant, $E_B$ is the binding energy of X$^{-}$, and $T$ is the temperature.  

According to equation 1, $n_e$ is directly linked to the ratio of $N_{X^{-}}$ to $N_{X^{0}}$. This relationship indicates that an increase in $n_e$ would result in a proportional rise in the relative concentration of X$^{-}$ compared to X$^{0}$. Note that the exciton concentration can be correlated with its PL intensity, such that $N_{X^{0}}$ $\sim$ $I_{X^{0}}$ and $N_{X^{-}}$ $\sim$ $I_{X^{-}}$, where $I_{X^{-}}$ and $I_{X^{0}}$ represent PL intensity of X$^{-}$ and X$^{0}$ respectively. Therefore, any change in electronic density within MoSe$_2$ should be manifested in the relative change in the population of X$^{0}$ and X$^{-}$, which would also lead to their respective PL emissions. 
\clearpage
}

\afterpage{%
\begin{figure*}
\includegraphics[width=\textwidth]{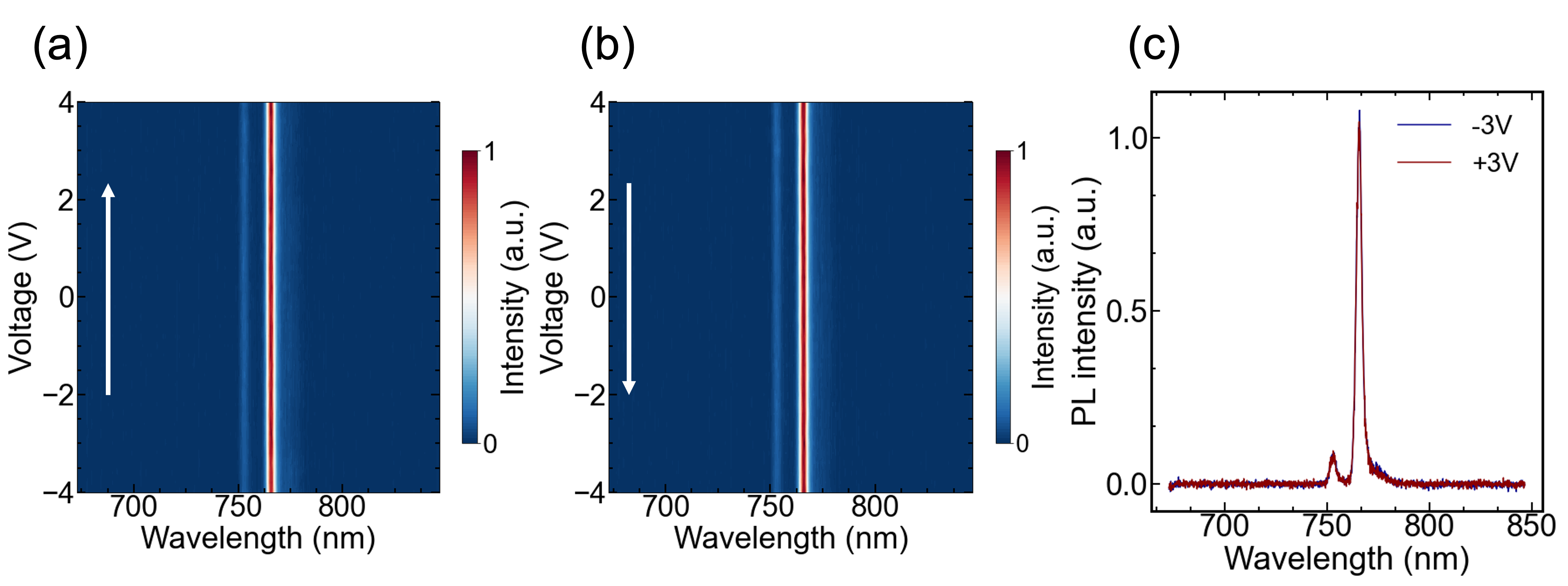}
\caption{\label{fig:wide} PL variation in a control device at 15~K. PL variation when (a) scanning Vpoling from $–4$~V to $+4$~V, (b)from $+4$~V to $–4$~V, (c) PL spectrum of monolayer MoSe$_2$ with $\pm$ 3~V pulse voltage modulation.} 
\end{figure*}
\clearpage
}

\afterpage{%
\begin{figure*}
\includegraphics[width=\textwidth]{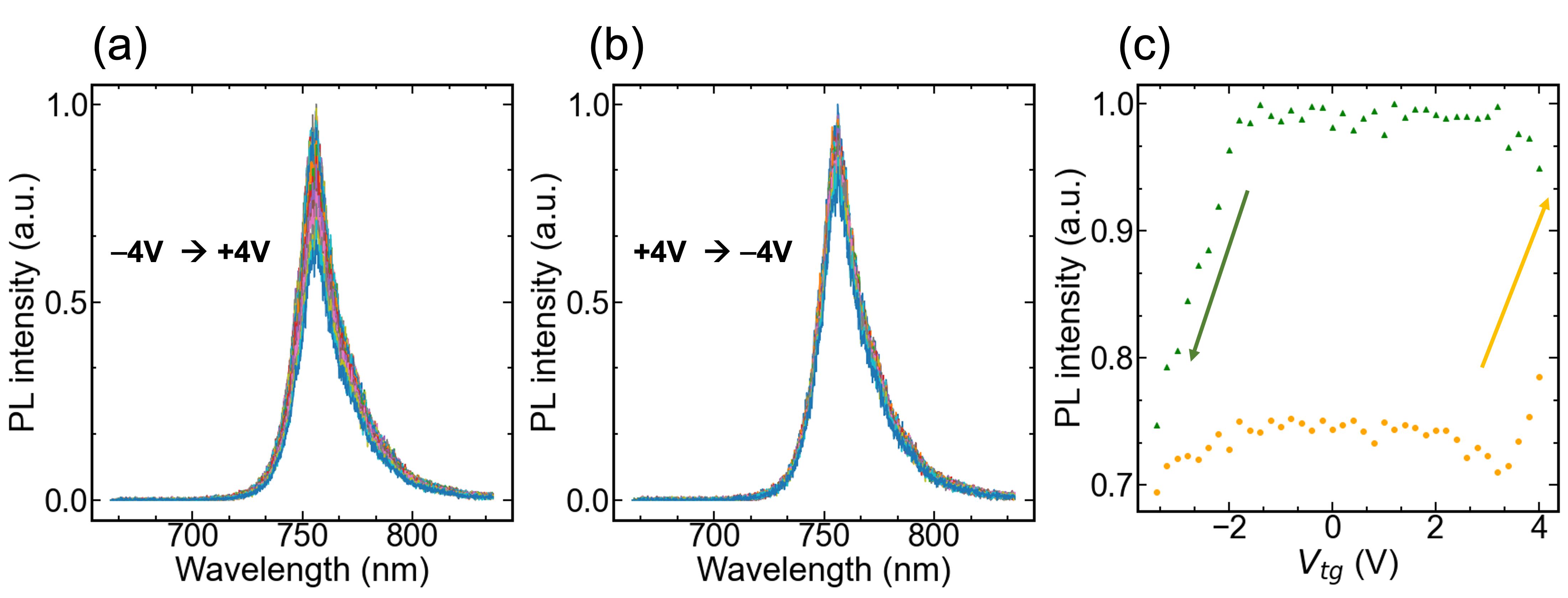}
\caption{\label{fig:wide} PL modulation at room temperature. Spectral change when (a) scanning Vpoling from $–4$~V to $+4$~V, (b)from $+4$~V to $–4$~V, (c) V$_{poling}$ dependent X$^{0}$ PL intensity shows a hysteresis loop.} 
\end{figure*}

At room temperature, PL spectrum is dominated by X$^0$ emission which is because thermal fluctuations induce electron escape from their trion bound state \cite{Mueller2018, Robert2016}. Therefore, the integrated PL intensity of spectra at room temperature can roughly correspond to X$^0$ emission. Its dependence on the poling voltage shows hysteretic behavior, which implies that the intensity ratio of X$^0$ and X$^{–}$ varies with the polarization state at room temperature. 
\clearpage
}

\afterpage{%
\begin{figure*}
\includegraphics[width=\textwidth]{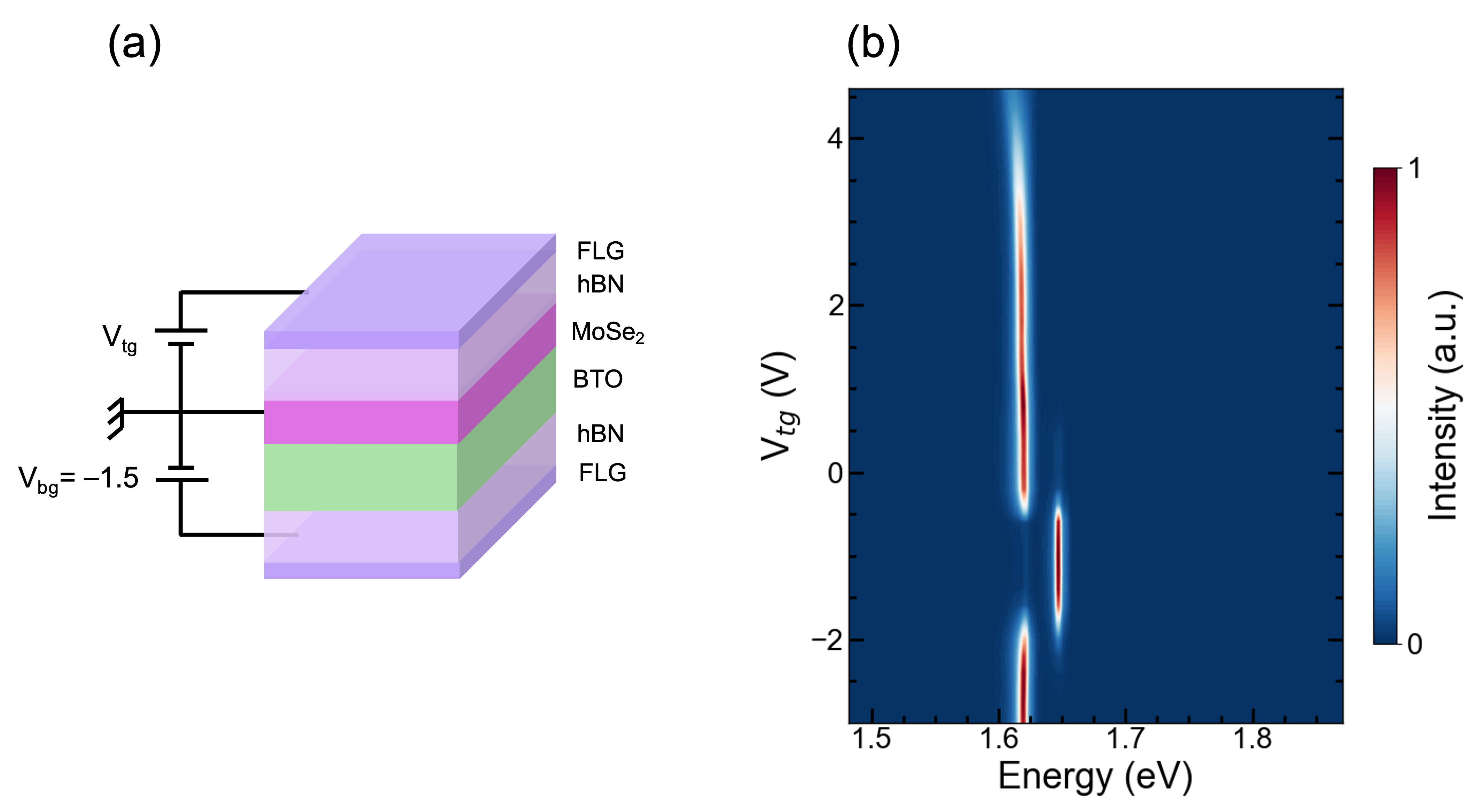}
\caption{\label{fig:wide} PL variation during the scanning of V$_{tg}$ at 10~K in DC mode. (a) We sweep the top gate (V$_{tg}$) from $–3$~V to $+4.2$~V in DC mode while maintaining the bottom gate at $–1.5$~V, (b) the resultant contour plot. It is notable that the charge neutrality point is observed to be approximately between –$2<V_{tg}<–0.6$.} 
\end{figure*}
\clearpage
}

\afterpage{%
\begin{figure*}
\includegraphics[width=\textwidth]{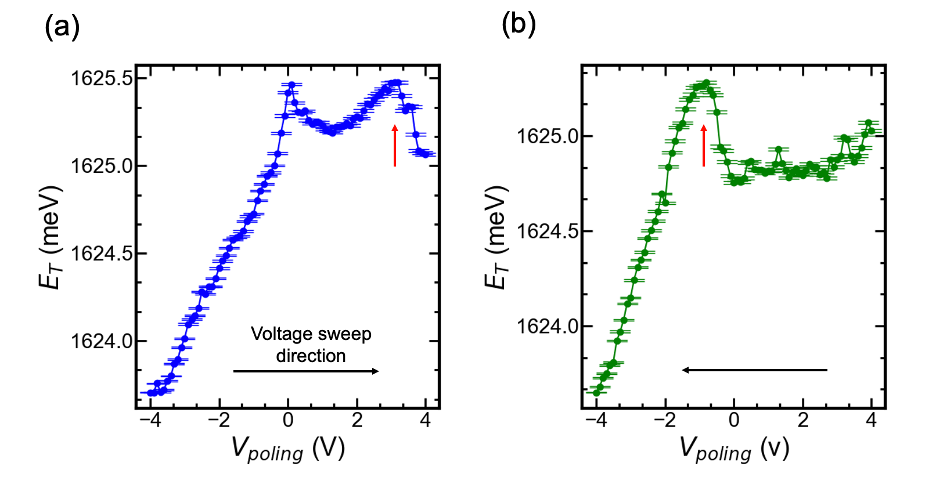}
\caption{\label{fig:wide} Trion energy (E$_T$) change in pulse mode. (a) E$_T$ change as we sweep Vpoling from $–4$~V to $+4$~V. There is an abrupt transition at V$_{poling}$=$+3$, which aligns with the hysteresis loop depicted in Fig.~2h in the main text, (b) E$_T$ alters as we sweep V$_{poling}$ from $+4$~V to $–4$~V.  The polarization switching occurs at V$_{poling}$ = $–0.8$ as shown in Fig.~2h, and there is a noticeable shift in E$_T$ when this switching happens. An additional bump is observed in both (a) and (b) at V$_{poling}$ $=0$, likely attributable to charge trapping in defects prevalent within MoSe$_2$ and BTO. Identifying other factors contributing to E$_T$ changes exceeds the scope of our study, given the complexity of the physics at a heterogeneous interface. Furthermore, variations in the ratio of PL intensity between X$^0$ and  X$^{-}$ serve as reliable indicators of changes in electron population, as phenomena like charge trapping do not exhibit dominant effect on this ratio. } 
\end{figure*}
\clearpage
}
\clearpage

\bibliography{Jaehong}